\documentclass[aps,prb,reprint,superscriptaddress,notitlepage]{revtex4-2}
\usepackage{times,amsmath,amsfonts,amssymb,mathrsfs,graphics,graphicx,color,comment,bm}
\usepackage[next]{inputenc}
\usepackage[dvips]{epsfig}
\usepackage{hyperref}
\usepackage{pdfsync}
\usepackage{appendix}
\usepackage{textcomp}
\usepackage{lipsum}  
\usepackage{gensymb}
\usepackage{graphicx}
\usepackage{upgreek}
\begin{document}

\title{$\mu$SR measurements of 3D Ising ferromagnetism in bulk Fe$_3$GeTe$_2$ crystals}

\author{M. N. Wilson}
\address{Memorial University, Department of Physics and Physical Oceanography, St. John's, Newfoundland A1B 3X7, Canada}
\author{J. Huang}
\address{Memorial University, Department of Physics and Physical Oceanography, St. John's, Newfoundland A1B 3X7, Canada}
\author{T. J. Hicken}
\address{PSI Center for Neutron and Muon Sciences, 5232 Villigen PSI, Switzerland}
\address{Durham University, Department of Physics, South Road, Durham, DH1 3LE, United Kingdom}
\author{D. A Mayoh}
\address{Department of Physics, University of Warwick, Coventry CV4 7AL, United Kingdom}
\author{B. M. Huddart}
\address{Clarendon Laboratory, Department of Physics, University of Oxford, Parks Road, OX1 3PU, United Kingdom}
\address{Durham University, Department of Physics, South Road, Durham, DH1 3LE, United Kingdom}
\author{G. Balakrishnan}
\address{Department of Physics, University of Warwick, Coventry CV4 7AL, United Kingdom}
\author{T. Lancaster}
\address{Durham University, Department of Physics, South Road, Durham, DH1 3LE, United Kingdom}

\begin{abstract}
Fe$_3$GeTe$_2$ is a quasi-two-dimensional itinerant ferromagnet with a van-der-Waals layered structure that exhibits substantial magnetic anisotropy. In bulk samples, this material becomes magnetically ordered below 220~K, while in thin samples, the gate-tunability of the ferromagnetic moment has attracted significant interest. Alongside high-temperature ferromagnetism, a low-temperature magnetic state occurs whose nature has not been conclusively identified, with some studies suggesting a transition between ferromagnetism and antiferromagnetism with temperature. Here we present magnetometry and $\mu$SR data that demonstrate that Fe$_3$GeTe$_2$ exhibits three-dimenional Ising-like ferromagnetism over the entire temperature range between $T_{C}=220$~K and  2~K. The feature at 170~K that has previously been proposed to be a transition to antiferromagnetism is instead likely to represent a domain-freezing transition.

\end{abstract}

\maketitle

\section{Introduction}

Spintronics seeks to  harness the degrees of freedom offered by electronic spin in devices, allowing high data storage density, novel computational paradigms, and reduced electricity use \cite{Zutic2004, Roy2024, Hirohata2020}. Advances in this field require magnetic materials whose properties can provide  control over the spin degrees of freedom . Particularly important  are atomically thin materials that could  be integrated into electronic devices at small length scales, achieving a high density of computational or storage elements. One promising avenue for the development of such materials is that of quasi two-dimensional (2D) systems that can be extracted from van-der-Waals layered magnetic crystals \cite{Zhang2024}. An example is the observation of long-range ferromagnetism near the 2D limit of atomic layers in Cr$_2$Ge$_2$Te$_6$, which   spurred  interest in such materials \cite{Gong2017}.

Fe$_3$GeTe$_2$ is an example of a van-der-Waals material that has been known for decades \cite{Abkrikosov1985}  and hosts ordered magnetic states \cite{Deiseroth2006}. Particular interest was motivated by the observation of tunable room-temperature ferromagnetism in extremely thin exfoliated films \cite{Deng2018}, which raises the possibility of device applications. In bulk form, the material is known to magnetically order around $T_{C}=220$~K \cite{Deiseroth2006}, with the reported presence of magnetic domains with short domain walls \cite{Birch2022,li2022} and large uniaxial anisotropies \cite{Deiseroth2006, wang2020}. However, the nature of this ordering has been debated, with variability between samples thought to arise from differing levels of Fe vacancies \cite{May2016}. Small changes in this Fe vacancy have been shown to significantly impact the magnetic ordering temperature and ordered moment \cite{Mayoh2021, Backes2024}, and careful crystal growth studies have shown that an excess of Fe in the starting materials is needed to produce stoichiometric crystals \cite{Mayoh2021}, suggesting that some behavior reported in the literature may unintentionally arise from composition variation. 

In addition, studies have identified Fe$_3$GeTe$_2$, with pure stoichiometry, as an unusual d-electron system that has heavy-fermion characteristics at low temperature \cite{Zhang2018, Vaidya2024}. There have also been  reports  of skyrmionic bubbles in this material under an applied field \cite{Ding2020,Birch2022,Powalla2023,Birch2024}. The character of these bubbles has also been shown to heavily vary depending on crystal stoichiometry \cite{Birch2024}, underscoring the importance of control of this parameter. More recently, magnon polarons have been observed which implying hybridization between the magnonic and phononic bands of excitations \cite{bansal2026}. These varied observed magnetic features highlight the complexity of the electronic states in this material.

Alongside these complex behaviors, there have been several reports of a second, lower-temperature magnetic transition in this material around $T=170$~K \cite{Yi2017,Chyczewski2023, Kim2019},  with suggestions that this involves a switch from ferromagnetism to antiferromagnetism with decreasing temperature \cite{Yi2017}. However, it has also been suggested that antiferromagnetic behavior order occurs only at the surface, caused by oxide impurities in the crystals \cite{Kim2019}. In this picture, the surface sensitivity of the measurement techniques employed amplify the antiferromagnetic signal that is otherwise absent in the bulk of the material. The differing claims make it important to further investigate the  bulk magnetism of Fe$_3$GeTe$_2$ to resolve the question of the magnetic state evolves at low $T$, and help inform how it may impact the complex behavior observed in this system.

In this paper, we use muon-spin relaxation ($\mu$SR), supported by magnetometry and density functional theory (DFT) determination of the muon sites, to investigate the bulk of high-quality stoichiometric crystals of Fe$_3$GeTe$_2$. Our work demonstrates that although two magnetic transitions are inherent to the bulk behavior of Fe$_3$GeTe$_2$, no change  between ferromagnetism and antiferromagnetism takes place as a function of temperature. Instead, we suggest that the changes in magnetism in this regime reflects anisotropic dynamics in the domain structure.  

\section{Methods}

Fe$_3$GeTe$_2$ crystals were grown using the chemical-vapor-transport method \cite{Mayoh2021}. With one of these crystals, orientation $H$~$||$ c-axis, we performed magnetometry measurements using a Quantum Design MPMS-XL SQUID magnetometer. Using the same sample, we also performed $\mu$SR measurements \cite{muontextbook2022}  at the Swiss Muon Source,  Paul Scherrer Institut, using the GPS spectrometer \cite{Amato2017} in zero applied magnetic field (ZF). For the $\mu$SR measurements we used 100~\textmu m of silver foil as a degrader, mounted in front of the sample to ensure a high fraction of muons stop inside the sample, with an expected muon penetration depth of approximately 200 $\mu$m. To aid interpretation of the $\mu$SR data, muon sites were calculated \cite{blundell2023_DFTmu} with the muFinder software package \cite{Huddart2022} that optimizes the crystal structure with a muon impurity via density functional theory (DFT) calculations using {\sc Castep} \cite{Clark2005}. Simulations of the muon spectra were then produced by calculating the dipole field at these muon sites using MuESR \cite{Bonifa2018}. 

\section{Results}

\begin{figure}[h]
\includegraphics[width=\columnwidth]{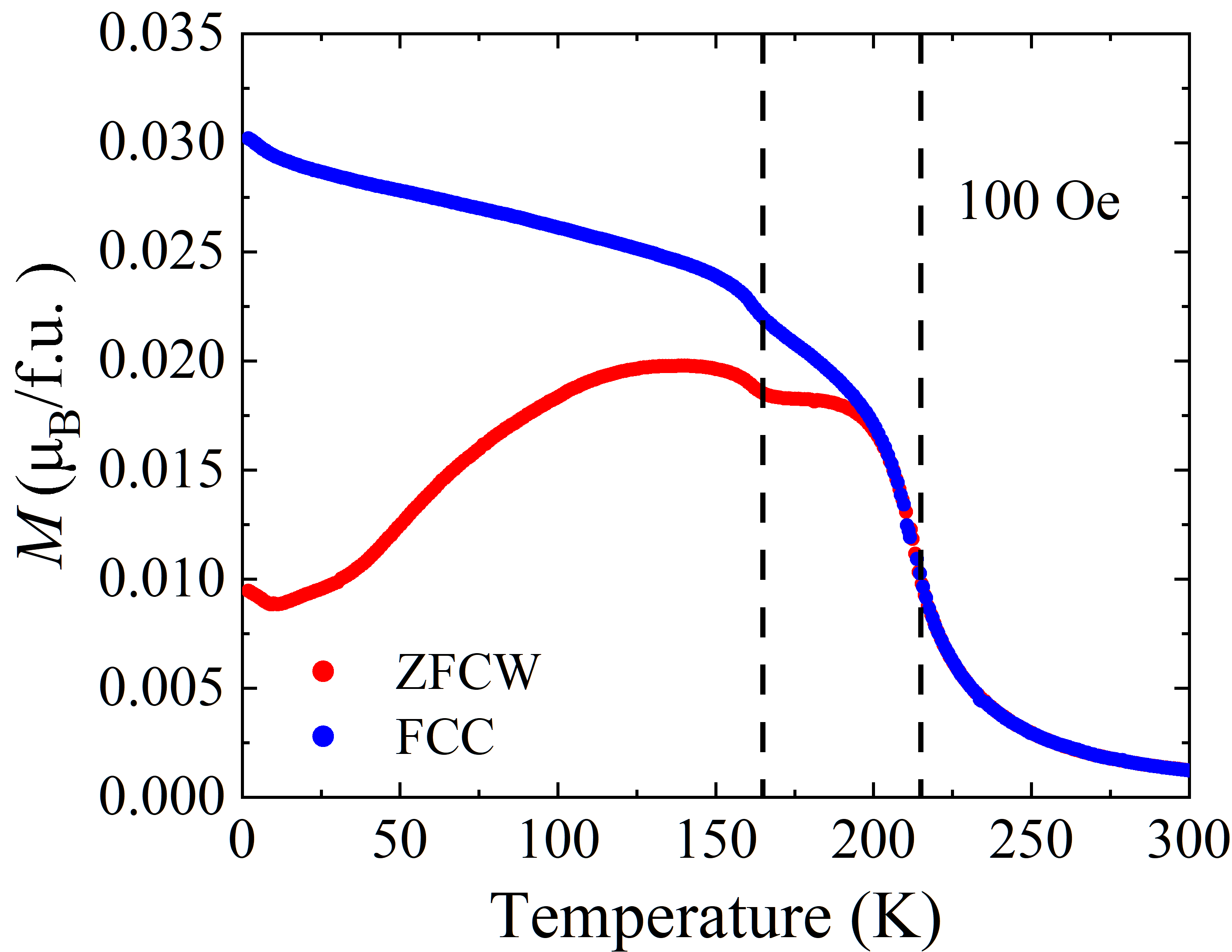}
\caption{Magnetization measurements of a Fe$_3$GeTe$_2$ crystal with a 100~Oe magnetic field oriented along the $c$-axis. Dashed lines indicate approximate temperatures of the magnetic transitions  (170~K and 220~K). Red data points are measured on warming after zero field cooling to 2~K, and blue points are measured while field cooling from 300~K.}
\label{fig:MvT}
\end{figure}

To characterize the magnetism in our samples, we first performed DC magnetic susceptibility measurements as a function of temperature with a magnetic field of $100$~Oe applied parallel to the $c$-axis, in order to identify the magnetic transitions. The results of these measurements are shown in Fig.~\ref{fig:MvT}; measured both on warming after zero-field cooling (ZFCW), and on cooling in a field (FCC). Both  curves suggest two magnetic transitions in the material associated with peaks in FCC $\frac{dM}{dT}$, one at approximately 220~K (corresponding to the transition from paramagnetism to ferromagnetic order), and one at around 170~K, below which the ZFCW susceptibility drops off, indicating the possible transition from ferromagnetism to a different magnetic state. The temperature of these transitions, and the shape of the magnetization curves, are broadly consistent with those of previous work (e.g.\ Ref.~\cite{Yi2017}). This suggests that measurements on our samples can be compared to those studies where an antiferromagnetic state was claimed. To identify the nature of these magnetic states, we now turn to $\mu$SR measurements.

\begin{figure}[t]
\includegraphics[width=0.7\columnwidth]{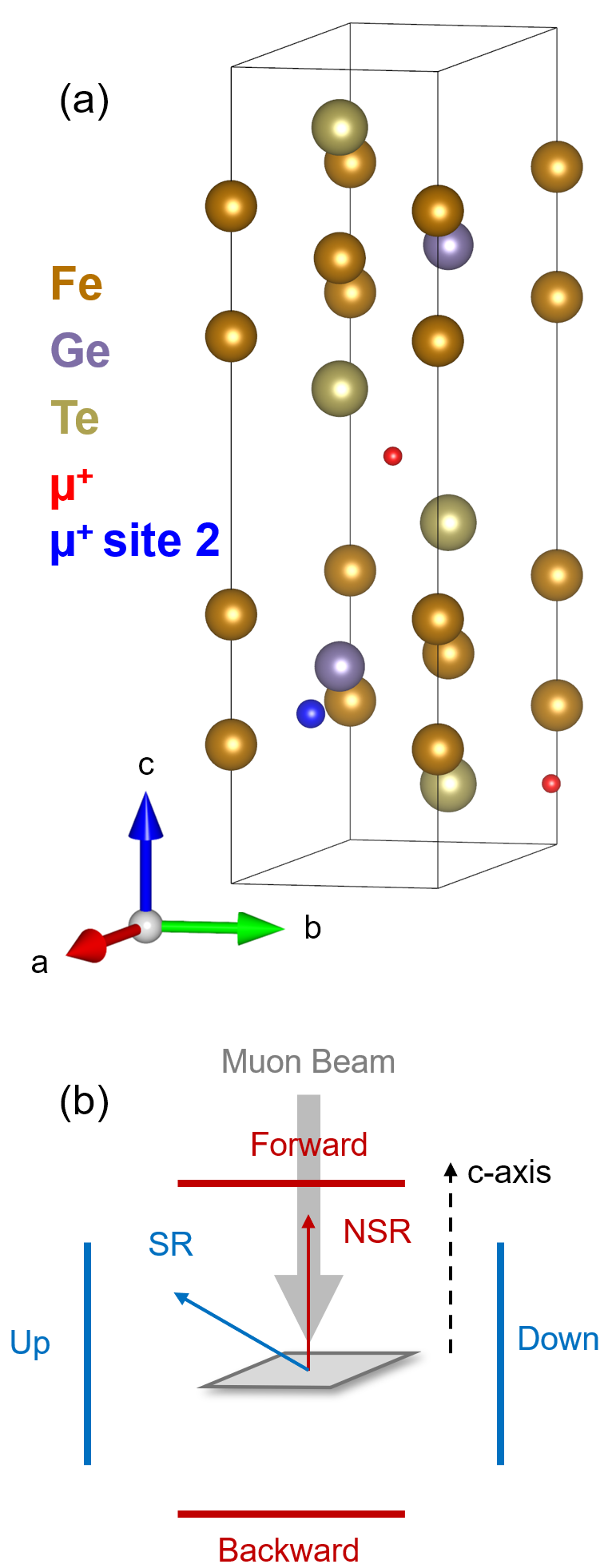}
\caption{(a) Crystal structure of Fe$_3$GeTe$_2$ illustrating the calculated muon sites (red; site 2 is shown as a larger blue sphere). (b) Geometry of the muon-spin relaxation measurements showing the initial muon-spin direction for non-spin-rotated (NSR) and spin-rotated (SR) modes, detector orientations, and the $c$-axis of the sample.}
\label{fig:MuonSite}
\end{figure}

Accurate interpretation of $\mu$SR data relies on knowledge of the position in which muons stop within a  crystal, in order to calculate the local magnetic fields experienced by muons in the different magnetic states. Our DFT calculations show the presence of three low-energy, symmetry-inequivalent muon sites, with positions given in Table~\ref{tab:muon} and shown as the red/blue spheres in Fig.~\ref{fig:MuonSite} (a). These calculations also suggest that the  muon does not cause a significant distortion to this material, indicating the $\mu$SR measurements should  probe the unperturbed magnetic state of Fe$_3$GeTe$_2$. 

\begin{table}[]
\begin{tabular}{|l|l|l|l|}
\hline
Site & Fractional Coordinates & Energy (eV) \\ \hline
1 & $(0.03 \pm 0.02, 0.99 \pm 0.02, 0.076 \pm 0.002)$      & $0 \pm 0.01$                    \\ \hline
2& $(0.40 \pm 0.01, 0.03 \pm 0.01, 0.176 \pm 0.003)$      & $0.254 \pm 0.001$                \\ \hline
3 & $(0.51 \pm 0.02, 0.50 \pm 0.02, 0.500 \pm 0.003)$      & $0.599 \pm 0.002$                \\ \hline
\end{tabular}
\caption{Candidate muon stopping sites listed in fractional coordinates for Fe$_3$GeTe$_2$, determined from DFT computations. Energies listed are the relative energies of the relaxed unit cell with an additional positive muon at that position, relative to the lowest-energy muon site.}
\label{tab:muon}
\end{table}

Using the information about the transition temperatures in our sample from magnetometry, we targeted $10$~K and $190$~K for initial zero-field $\mu$SR measurements in the two different magnetic states. We performed measurements with the $c$-axis of our crystal parallel to the incoming muon beam, and measured with incoming muon spin in two polarizations: (i) parallel to the $c$-axis measuring polarization in the forward-backward detector pair [non-spin-rotated (NSR) mode], and (ii) with the muon spin polarization rotated approximately 60$^{\circ}$ away from the $c$-axis [spin-rotated (SR) mode] where we can look at either the polarization parallel to $c$ (using the forward-back detectors) or perpendicular to $c$ (using the up-down detector set). This geometry is shown in Fig.~\ref{fig:MuonSite}(b). It is important to note that the 200~$\mu$m penetration depth of the muons in the sample means that, on an atomic scale, we are probing the bulk behavior,  far away from the influence of surface oxides that could affect the results of other surface-sensitive measurements. 

\begin{figure}[t]
\includegraphics[width=\columnwidth]{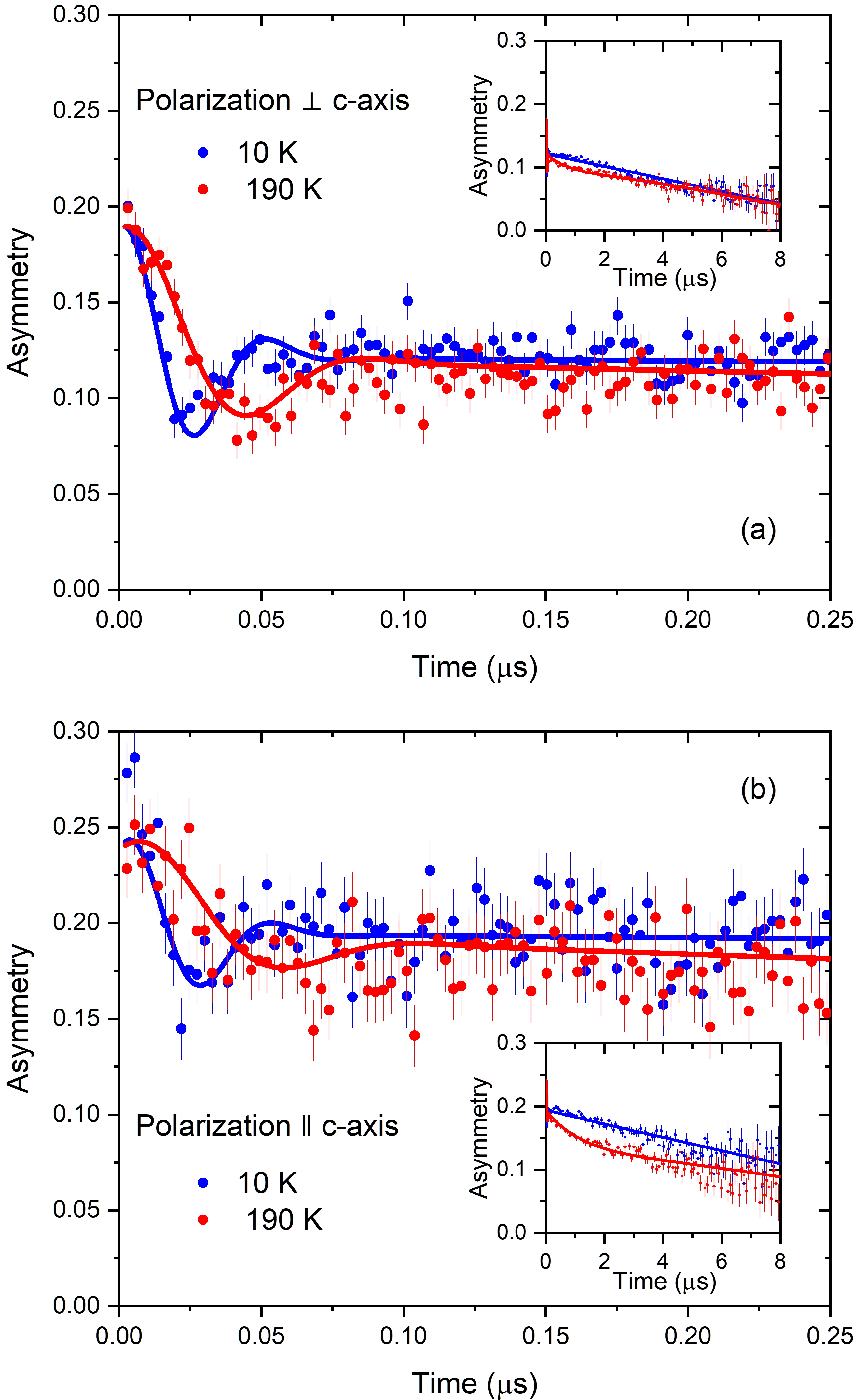}
\caption{Example ZF $\mu$SR data  Fe$_3$GeTe$_2$, shown at 10~K (blue) and 190~K (red) measuring  (a)  polarization perpendicular to the $c$-direction (SR mode), and (b)  polarization parallel to the $c$-direction of (NSR mode). Solid lines show fits to Eq.~\ref{eq:muonFit}. Insets to each panel show the same data over a longer time range (0 - 8~$\mu$s), highlighting the  slow relaxation of the signal.}
\label{fig:10-190}
\end{figure}

The results of measurements of the muon polarization perpendicular to the $c$-direction (measured in SR mode) are shown in Fig.~\ref{fig:10-190}(a). These show a rapidly damped oscillation with a slowly relaxing tail seen in the inset. Figure~\ref{fig:10-190}(b) shows the polarization parallel to $c$ (measured in NSR mode) and displays similar behavior. 
The spectra we measure are consistent with a previous report of the early-time behavior seen in a weak transverse field $\mu$SR measurement on this material \cite{drachuck2018}.
The observation of such a rapidly-relaxing signal, rather than long lived oscillations, indicates the presence of significant static disorder in the magnetic structure. One place where this could arise is stacking faults, as have been previously observed in this material \cite{Park2022}. Such stacking faults lead to variations in the interlayer magnetic coupling strength, and could cause substantial spatial variation of the magnetic moment. This would lead to a wide distribution of internal magnetic field strengths in the material, resulting in different muons precessing at slightly different rates and hence to a rapidly relaxing signal. 

The most important feature of our data is that the signal is qualitatively unchanged between 190~K and 10~K for both polarization directions. At each of these temperatures, the data is characterized by the same sort of relaxing oscillation with consistent amplitude, and a single frequency that is approximately 1.5 times higher at lower temperatures. (This increase in frequency is consistent with an increase in the ordered magnetic moment  at lower temperature arising from the usual thermal effects in an ordered magnet.) To better understand how the observed behavior compares to that expected for the antiferromagnetic and ferromagnetic states, we now turn to simulations of the $\mu$SR spectra using our calculated muon sites.

\begin{figure}[t]
\includegraphics[width=\columnwidth]{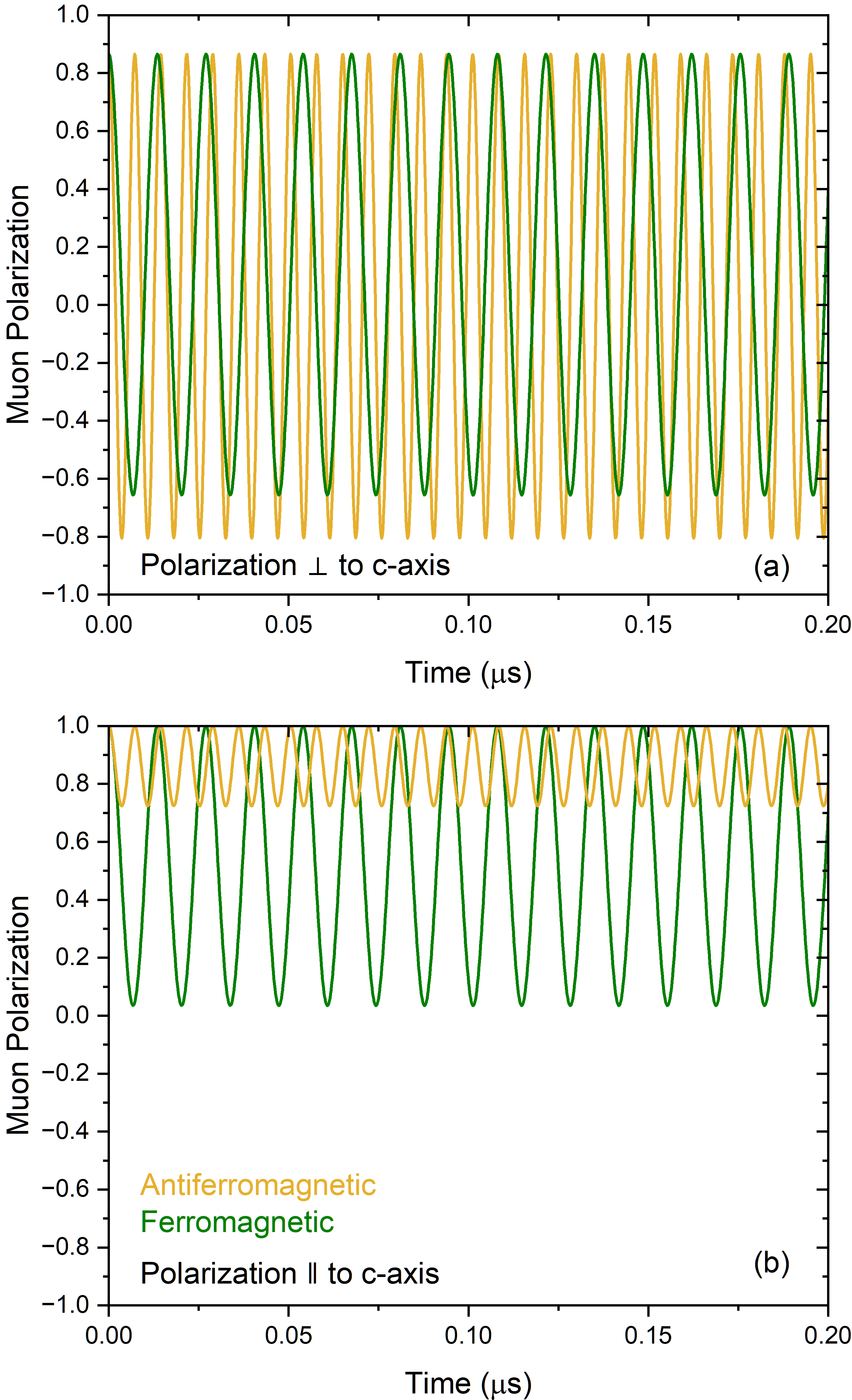}
\caption{Simulated spectra for ferromagnetic and antiferromagnetic magnetic structures of Fe$_3$GeTe$_2$, calculated for site 2 from Table~\ref{tab:muon}. (a) Muon-spin polarization perpendicular to the $c$-axis (parallel to $a$-axis). (b) Muon-spin polarization parallel to the $c$-axis.}
\label{fig:SimSpec}
\end{figure}

The two magnetic structures previously proposed for Fe$_3$GeTe$_2$ at different temperatures are (i) a ferromagnetic state (1.8 $\mu_B$ / Fe moments parallel to the $c$-axis), and (ii) an antiferromagnetic state with neighboring layers of Fe atoms anti-aligned  (1.8 $\mu_B$ parallel or anti-parallel to $c$) \cite{Yi2017}. To distinguish the expected behavior from these two states we simulated their muon-spin relaxation spectra, assuming no disorder-induced damping. We found that the most straightforward picture consistent with our measurements requires that only site 2 is occupied (although other, more complicated, combinations of muon sites also lead to similar conclusions: see the Appendix for a discussion). The result of our simulations is shown in Fig.~\ref{fig:SimSpec} for muon-spin polarization parallel and perpendicular to the $c$-axis. The simulations predict different behavior for the two magnetic structures. For the antiferromagnetic state the simulations yield almost no precessing amplitude for muon polarization parallel to $c$, while for the ferromagnetic state the precession amplitude is slightly suppressed compared to the perpendicular polarization, but over half of the precessing amplitude remains. The simulations also show that, at the muon site, there are large magnetic field components parallel to the $c$-axis for both the ferromagnetic and antiferromagnetic cases, but that the fields for these two states differ by almost a factor of two. This shows up in the predicted spectra as robust precession with similar amplitude for muons with spin  perpendicular to the $c$-axis in both magnetic states, but a much more rapid oscillation for the antiferromagnetic state.

There are two notable differences between the simulations and the data. First, these simulations present an idealized picture where there is no disorder or dynamic component to the muon signal. The first of these effects would result in damping of the oscillations, while the second would introduce a relaxing baseline muon polarization; both effects are seen in our data. Phenomenological damping can be added to the simulations to account for these effects to match the damping seen in the data. Second, even for the lower field seen in the ferromagnetic state, the oscillation rate in the simulations is substantially higher than that seen in the data. This would suggest that the ordered magnetic moment in these samples at 10~K is substantially lower than the 1.8~$\mu_B$ / Fe that we had assumed. Specifically, to match our observed oscillation frequency, the moment would need to be 0.45~$\mu_B$ / Fe. 

\begin{figure}[b]
\includegraphics[width=\columnwidth]{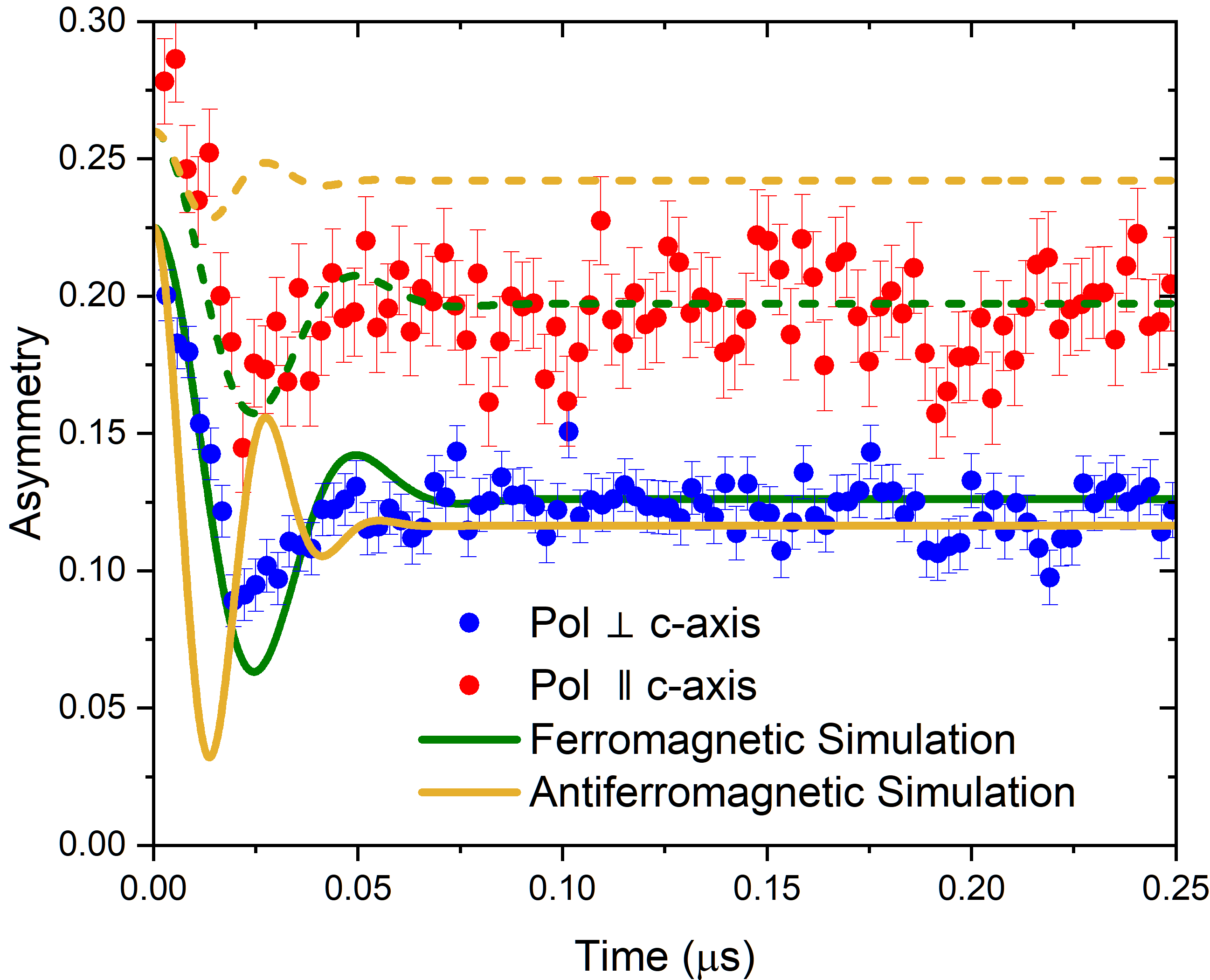}
\caption{Muon-spin relaxation data shown at 10~K, measuring the polarization parallel to $c$ in NSR mode  (red), and polarization perpendicular to $c$ in SR mode  (blue). Curves show the simulated $\mu$SR spectra for the ferromagnetic (green) and antiferromagnetic (yellow) states. Dashed lines show simulations for polarization parallel to c-axis, solid lines for polarization perpendicular to c. }
\label{fig:muonDat}
\end{figure}

However, even with these differences, the simulations demonstrate that muon-spin relaxation measurements can qualitatively distinguish between the two possible magnetic states. From the simulations, we would expect the measurements to show a substantial difference between the two polarization directions if the sample is in an antiferromagnetic state (notably, a very large change in precession amplitude), but more similar behavior if the sample is in a ferromagnetic state. Figure \ref{fig:muonDat} presents a direct comparison of the 10~K $\mu$SR data with the simulations for both states. In these simulations, we have added in phenomenological damping to match the data (gaussian damping with $\sigma = 37\mu$s$^{-1}$), used the lower 0.45~$\mu_B$ / Fe moment value, and added in a 50\% background asymmetry (slow relaxing, $\sigma_{bkg} = 0.09\mu$s$^{-1}$) in both cases to match our data. The simulation for polarization parallel to c is scaled by an initial asymmetry of 0.26, while that for polarization perpendicular to c is scaled by $0.26 \sin 60 = 0.225$ to account for the reduced muon polarization in this direction for the spin-rotated mode used to measure this data set. This comparison shows that the oscillation amplitude differences between polarization directions we see in the data agree well with the ferromagnetic case, but are inconsistent with antiferromagnetism.

In addition, the simulations suggest that we would expect to see a large increase in the measured internal field if we crossed over a transition from ferromagnetism to antiferromagnetism as a function of decreasing temperature. To investigate this possibility, we now turn to a more thorough investigation of the temperature dependence of the $\mu$SR data. We measured spectra as a function of temperature between 10~K and 250~K in both polarization directions. The data at all temperatures and both incident muon polarizations is parameterized using the equation,
\begin{equation}
A(t) = A_1 \cos (\gamma_\mu Bt + \phi )e^{-\frac{\sigma_1 t^2}{2}} + A_2 e^{-\lambda t} + A_3 e^{-\frac{\sigma_2 t^2}{2}},
\label{eq:muonFit}
\end{equation}
where $A$ is the total asymmetry, $A_1$ is the oscillating asymmetry, $A_2$ is the asymmetry of an exponentially damped term that describes dynamics, $A_3$ is a background asymmetry, $\phi$ is the phase of the oscillating term, $\gamma_\mu$ is the muon gyromagnetic ratio, $B$ is the internal field, $\sigma_1$ is the damping rate of the internal field, $\lambda$ is a dynamic relaxation rate, and $\sigma_2$ is a small background damping rate. In this model, $\phi$, $\sigma_2$, and the asymmetries are temperature independent, $B$ and $\lambda$ are allowed to vary as a function of temperature, and $\sigma_1$ scales with $B$ ($\sigma_1 = 37\mu$s$^{-1}$ at 10~K). We find that the data at all temperatures and both polarization directions are well described by this model, with no need to introduce additional components, change the model, or vary the relative size of the asymmetries crossing over the observed secondary magnetic transition at 170~K, as might be expected if the structure of the local magnetism substantially changed.

\begin{table}[b]
\begin{tabular}{|l|l|l|l|}
\hline
Parameter & NSR FB & SR FB & SR UD \\ \hline
$A_1$ & $0.049 \pm 0.002$ & $0.029 \pm 0.001$ & $0.068 \pm 0.001$                 \\ \hline
$A_2$ & $0.074 \pm 0.001$ & $0.042 \pm 0.001$ & $0.032 \pm 0.01$                  \\ \hline
$A_3$ & $0.121 \pm 0.001$ & $0.066 \pm 0.01$ &   $0.089 \pm 0.01$                \\ \hline
$\sigma_2$   & $0.08 \pm 0.01$~$\mu$s$^{-1}$ & $0.10 \pm 0.01$~$\mu$s$^{-1}$ & $0.09 \pm 0.01$~$\mu$s$^{-1}$                \\ \hline

\end{tabular}
\caption{Temperature dependent fitting parameters.}
\label{tab:TInd}
\end{table}

\begin{figure}[t]
\includegraphics[width=0.95\columnwidth]{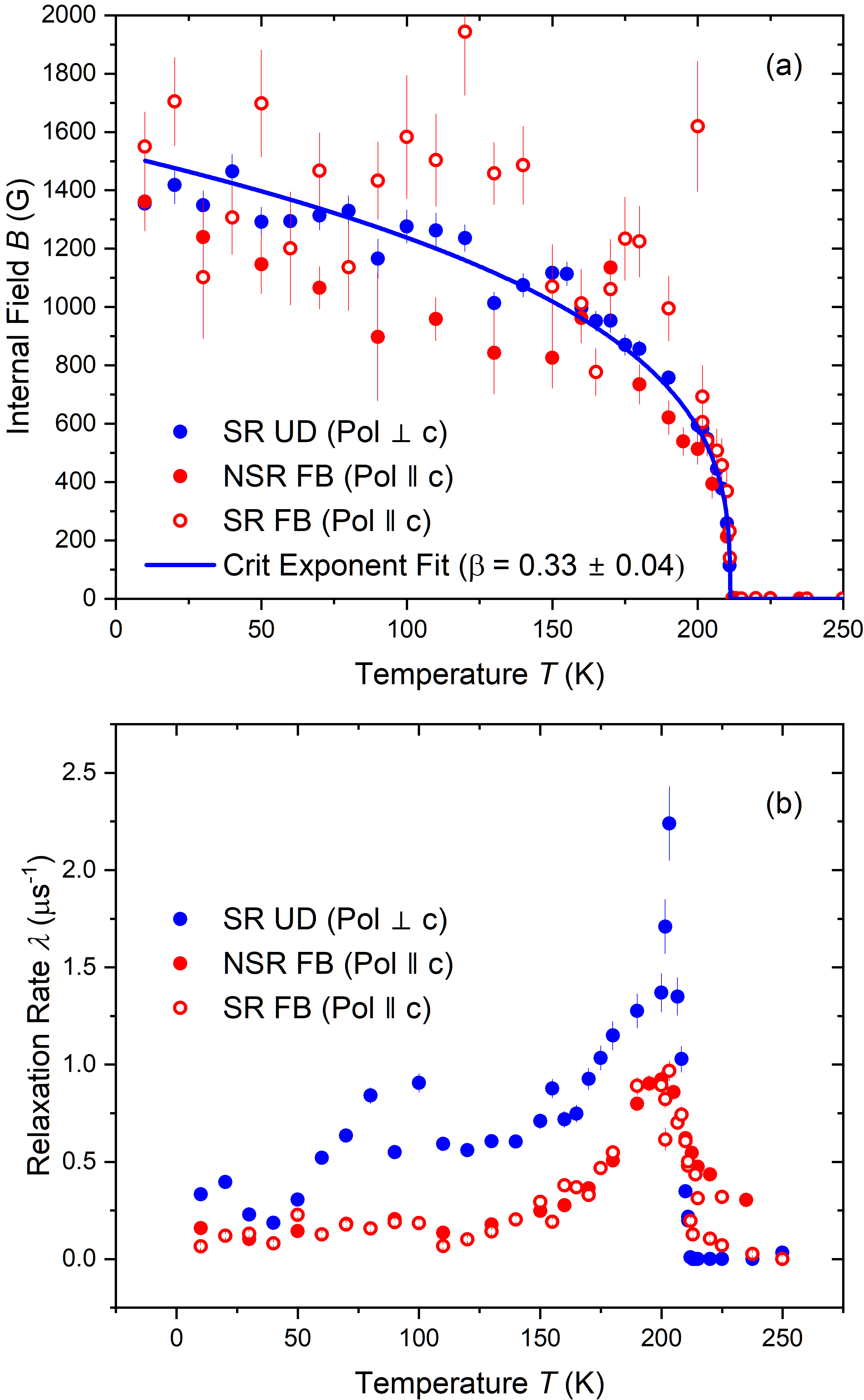}
\caption{$\mu$SR fit parameters resulting from fitting the data from Fig.~\ref{fig:muonDat} (and similar data at other temperatures) to Eq.~\ref{eq:muonFit}. Solid line in panel (a) shows a fit to 3D Ising critical behavior with Eq.~\ref{eq:Critical}.}
\label{fig:FitParam}
\end{figure}

Table \ref{tab:TInd} shows the values for the temperature-independent parameters for all three measurement modes (NSR with forward-back detectors, SR with up-down detectors, and SR with forward-back detectors), while Fig.~\ref{fig:FitParam} shows the temperature dependent parameters. Notably, panel (a) shows the temperature dependence of the internal field which is consistent for the fitting of all three data sets. This shows a smooth temperature dependence for both polarizations from the upper transition temperature at 220~K down to low temperatures. Crossing the putative 170~K transition, the temperature dependence remains smooth, with no sudden change. This temperature dependence can be well fit with power law behavior following, 
\begin{equation}
B(T) = A \left(\frac{T_C-T}{T_C}\right)^{\beta},
\label{eq:Critical}
\end{equation}
where $A$ is a constant, $T_C$ is the transition temperature, and $\beta$ is the critical exponent. Specifically, we note that an exponent of $\beta = 0.33 \pm 0.04$, consistent with 3D Ising magnetism previously seen in bulk Fe$_3$GeTe$_2$ \cite{Fei2018}, results in an extremely good fit throughout the whole temperature range, including close to the transition, where we expect critical behavior. The upper transition temperature fits to $T_C = 211.1\pm 0.1$~K, slightly lower than the transition seen in magnetometry, but consistent with differences commonly seen between different experimental techniques. This consistent temperature dependence of the internal field strength across the observed lower transition and in both muon polarization modes provides strong additional evidence that the magnetism in Fe$_3$GeTe$_2$ does not change  between ferromagnetism and antiferromagnetism as a function of temperature. As noted previously, if it did change to antiferromagnetism upon cooling, we would have instead observed a large increase in the internal field cooling through the transition, inconsistent with our data.

Figure~\ref{fig:FitParam}(b) shows the temperature dependence of the dynamic relaxation rate $\lambda$. These data show relaxation rates that are similar for the two muon polarizations at low temperatures, with a broad peak between the 170~K and 220~K regions that is much larger in magnitude for the perpendicular polarization. The width of this peak in temperature is substantially larger than would be expected of critical behavior across a typical second-order phase transition, and therefore likely some of the dynamics in this temperature region has a different origin.

$\mu$SR is sensitive to dynamics in the magnetization for fluctuations in the local magnetic field perpendicular to the initial muon-spin polarization direction. These data therefore suggest that dynamics are fairly isotropic at the lowest temperatures but gain  anisotropy between the two magnetic transitions, with enhanced dynamics with fluctuating amplitude in the $c$-direction. One way in which this could happen is if a rapidly-fluctuating domain structure exists between 170~K and 220~K, with ferromagnetic domain boundaries existing within the $a$-$b$ plane \cite{Birch2022,li2022}. Thermal fluctuations of these boundaries on the MHz time scale would persist over a larger temperature range than typical critical dynamics, explaining the temperature width of the relaxation rate peak. They would also be inherently anisotropic, explaining the larger dynamic relaxation rate observed for the polarization perpendicular to $c$. In this picture, the bulk transition at 170~K would be a domain-freezing transition, leading to a state with minimal dynamics on the muon timescale, and an overall compensated net magnetization that is expected to approach zero (leading to the drop in the total net magnetic moment of the sample seen in the magnetometry data at low temperature, and explaining previous identification of this state as antiferromagnetic). This interpretation would be consistent with the data we have presented here, without the need for the antiferromagnetic transition that has been invoked in previous work.

\section{Conclusion}
We have presented magnetometry and $\mu$SR data, backed by DFT calculations of muon stopping sites, and simulations of expected $\mu$SR spectra to investigate the magnetism in Fe$_3$GeTe$_2$. Taken as a whole, this investigation suggests that the material exhibits 3D Ising-like ferromagnetism over the entire temperature range between its Curie temperature of 220~K and our base temperature of 2~K. The transition at 170~K that has previously been proposed to be a transition to antiferromagnetism is instead identified to likely be a domain-freezing transition, leading to a low-temperature state which is still ferromagnetic. 

\section{Appendix}

Identification of the volume fraction of muons stopping at different candidate muon sites is a continuing challenge for $\mu$SR measurements and simulations \cite{blundell2023_DFTmu}. In our case, we have three candidate muon sites as indicated in Table~\ref{tab:muon}. The first two of these sites are relatively similar in energy (spaced by only 0.25~eV), suggesting that either or both are reasonably likely to be occupied. Site 3, on the other hand, is 0.6~eV higher in energy than site 1. Energy differences on this scale  typically mean that there is a low probability that this site will be occupied with any substantial volume fraction. We therefore exclude site 3 from our analysis and focus on sites 1 and 2.

\begin{figure}[t]
\includegraphics[width=\columnwidth]{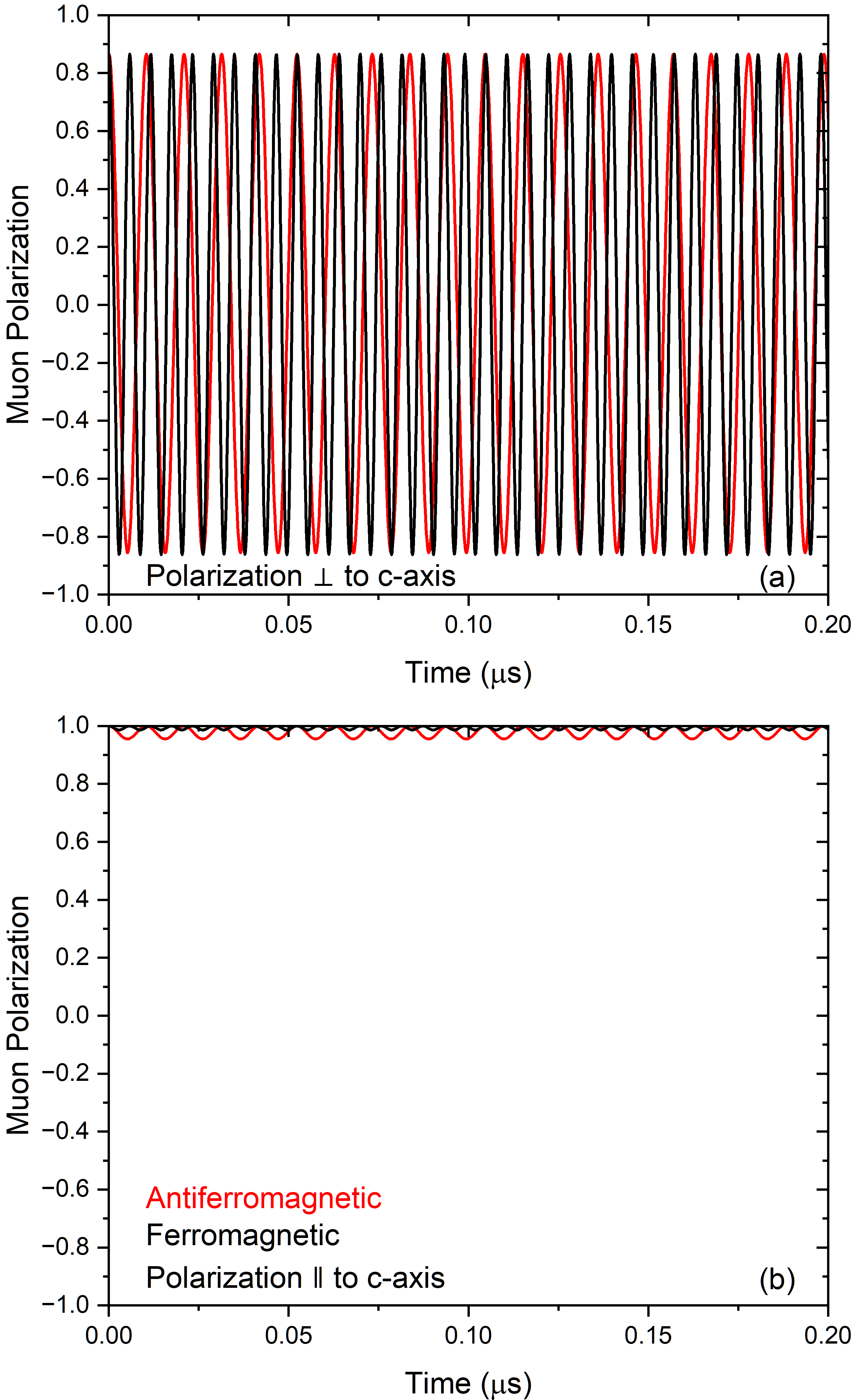}
\caption{Simulated spectra for ferromagnetic and antiferromagnetic states of Fe$_3$GeTe$_2$ calculated using site 1 from Table~\ref{tab:muon}. (a) shows the simulation for muon spin polarization perpendicular to the $c$-axis (parallel to $a$-axis), while (b) shows the simulation for muon spin polarization parallel to the $c$-axis.}
\label{fig:SimSpecS1}
\end{figure}

For these two sites, the calculated spectra are shown in Fig.~\ref{fig:SimSpec} (site 2) and Fig.~\ref{fig:SimSpecS1} (site 1). The striking difference that we immediately see from these simulations is that site 1 shows nearly-zero oscillating amplitude for both magnetic states with a muon polarization initially parallel to the $c$-axis, while site 2 shows large suppression of the oscillations in this polarization only for the antiferromagnetic state. In our data, we see significant oscillations present for both muon polarizations (inconsistent with either state for site 1). This suggests to us that site 1 is likely not occupied, or at least is not the dominant muon site. While the lowest energy site is often the one that is seen to be occupied in experiments, factors such as the relative size of energy barriers for muons to hop out of these sites can influence this and cause a different site to be occupied, as is suggested by our data in this work. 

\begin{figure}[t]
\includegraphics[width=\columnwidth]{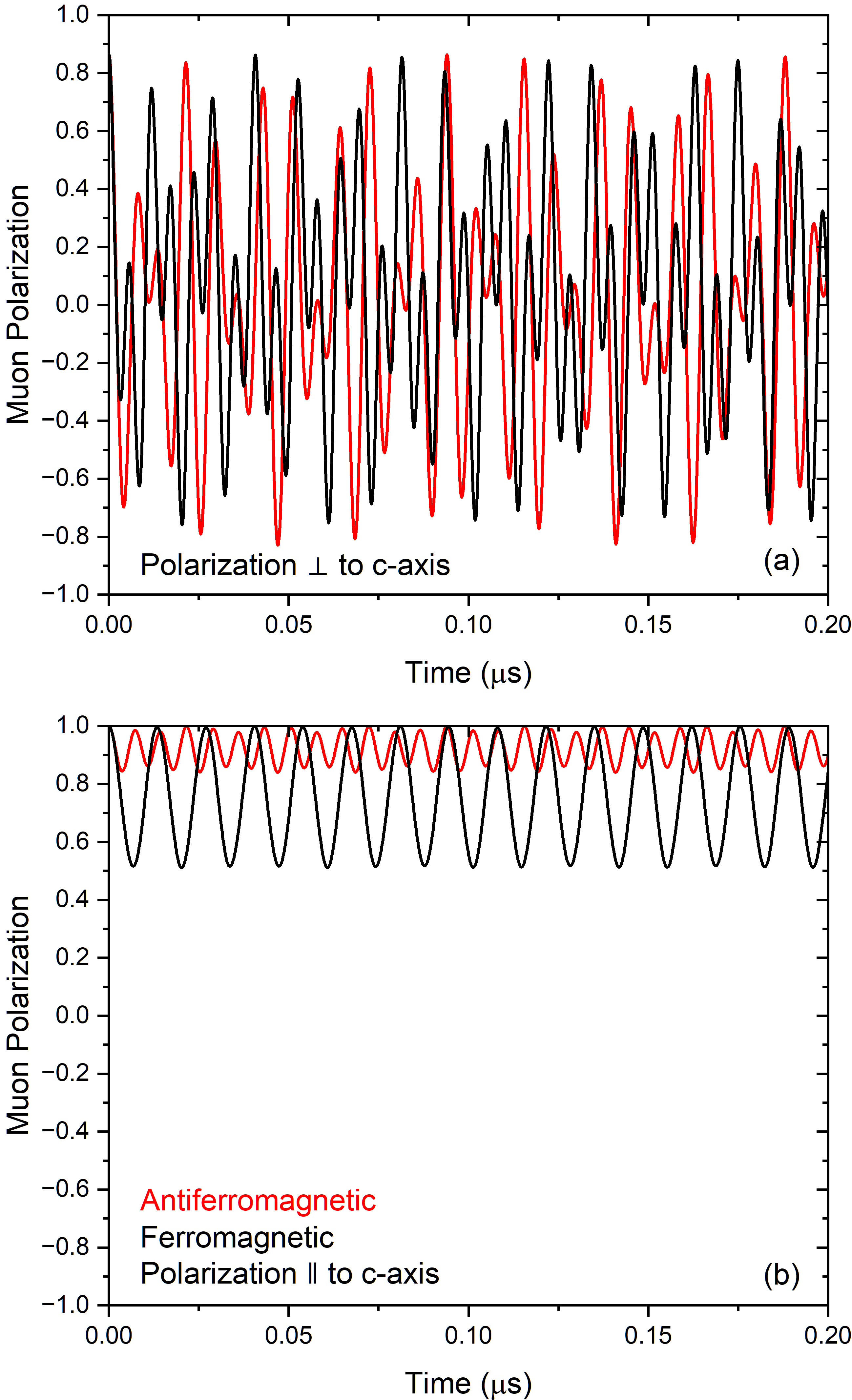}
\caption{Simulated spectra for ferromagnetic and antiferromagnetic states of Fe$_3$GeTe$_2$ calculated using site 1 and 2 from Table~\ref{tab:muon}. (a) shows the simulation for muon spin polarization perpendicular to the c-axis (parallel to $a$-axis), while (b) shows the simulation for muon spin polarization parallel to the $c$-axis.}
\label{fig:SimSpecS12}
\end{figure}

For completeness, we present in Fig.~\ref{fig:SimSpecS12} simulations assuming a 50\% occupation of sites 1 and 2. These plots show a qualitatively similar picture to that of site 2 alone; the oscillating amplitude is much more dramatically suppressed in the antiferromagnetic state for parallel polarization, both states show similar amplitude oscillations in the perpendicular polarization, and a transition between the two states would be expected to show a significant change in the oscillation rate. The main qualitative difference is that the perpendicular polarization graph shows two oscillation frequencies beating against each other, while the site 2 data and our fits show only a single frequency. This indicates that our data is more consistent with only site 2 being occupied, and hence we focus the discussion of the main manuscript on simulations with 100\% muon fraction in site 2, for simplicity.

It should be noted that the high damping rate of our observed signal could obscure the presence of a second oscillating frequency. If this were present, it might suggest that we do have some fraction of the muons stopping in site 1. However, the qualitative conclusions of our work would remain unchanged if we assumed that we had 50/50 occupation of site 1 and site 2, instead of 100\% in site 2. 

\begin{acknowledgements}
Some of the measurements presented here were made at the Swiss Muon Source, Paul Scherrer Institute, and we are grateful for provision of beamtime. 
This work was supported by the UK Skyrmion Project EPSRC (UK) Programme Grant (EP/N032128/1). M.~N.~Wilson acknowledges the support of the Natural Sciences and Engineering Research Council of Canada (NSERC). G.B., D.A.M., and T.L acknowledge EPSRC grant EP/Z534067/1. G.B. and D.A.M. acknowledge EPSRC grant EP/Z535874/1. Research data presented here will be made available via DOI:XXX.
\end{acknowledgements}

\end{document}